\documentclass[12pt]{iopart}

\usepackage{iopams}
\usepackage{setstack}
\usepackage{amssymb}
\usepackage{graphicx}
\usepackage{tabularx,caption}
\usepackage{wrapfig}
\usepackage{subfig}
\usepackage{float}
\usepackage{paralist}
\usepackage{algorithmic, algorithm}
\usepackage{multicol}
\usepackage{dsfont}

\begin{document}

\title[Interpolation of Microscale Stress and Strain]{Interpolation of Microscale Stress and Strain Fields Based on Mechanical Models}

\author{Wenzhe Shan*, Udo Nackenhorst}

\address{Institute for Advanced Studies, Nanchang University, Jiangxi, China}
\ead{shan@ncu.edu.cn}
\begin{abstract}
In this short contribution we introduce a new procedure to recover the stress and strain fields for particle systems by mechanical models. Numerical tests for simple loading conditions have shown an excellent match between the estimated values and the reference values. The estimated stress field is also consistent with the so called Quasicontinuum stress field, which suggests its potential application for scale bridging techniques. The estimated stress fields for complicated loading conditions such as defect and indentation are also demonstrated.
\end{abstract}

\maketitle

\section{Introduction}
Multiscale methods have been attracting more and more attention for the last two decades. One of the main applications of the multiscale methods is to study material with defects \cite{Ortiz2001, Shilkrot2004, Gumbsch1995, Abraham2000}. The region containing defects cannot be accurately described by the traditional continuum mechanics and atomic models have been used as a powerful tool to study the defects at small scale. However the high computational cost makes it difficult, if not impossible, to use a fully atomic model for problems of sizes large enough to be of any practical applications, and therefore multiscale methods which can couple the molecular dynamics and the continuum mechanics are called for. Good review papers on such methods can be found at \cite{Fish2006, Nieminen2002, Muralidharan2003, Vvedensky2004, Cherkaoui2009}. Among the existing results, we consider the Quasicontinuum method \cite{Tadmor1996a, Tadmor1996b, Miller2002} as one of the most systematically developed multiscale framework. It introduces a physically consistent coupling between the continuum mechanics and the molecular dynamics. Though showing promising results in the material defect analysis at nanoscale \cite{Tadmor1996a, Tadmor1999, Miller1998, Dupont2008, Dobson2007}, there are still some problems to fix and many extensions to make, and one of them is to develop a method to interpolate the stress and strain fields for the microscale region of Quasicontinuum models. Stress and strain are only defined in continuum mechanics, therefore, without extra interpolation techniques, one cannot obtain the stress and strain fields for the microscale region which is modeled by molecular dynamics. On the other hand, the microscale stress fields are also found useful for studying the material behaviors at the microscale \cite{Egami1980, Hull2001, Ray1984}. Moreover, the microscale stress itself has recently been exploited as part of the coupling variables \cite{Fish2007} for obtaining smoother transition at the macro-micro boundary. Therefore, it is desirable to develop techniques that can interpolate the stress and strain fields for the particle methods which are consistent with their counterparts in continuum mechanics.

Various methods for obtaining the microscale stress tensor have already been available in literature. Among them, the so called Virial stress \cite{Irving1950} provides simple expressions for the atomics stress and is still widely used in the literature, despite the controversial debate over its validity \cite{Zhou2003, Subramaniyan2008, Leo_unpub}. An appropriate version was put forward by Lutsko \cite{Lutsko1988} and later improved by Cormier \cite{Cormier2001}, who derived it in a more general form. However, the derivation process is found to be quite complicated. Fourier transformations and inverse transformations have been applied to solve the differential equations \cite{Lutsko1988, Cormier2001, Zhou2003}, without proving their applicability. A non-invertible tensor expansion which is mathematically true but physically weak must be applied during the derivation \cite{Cormier2001, Zhou2003}. The oscillating effect of the interpolated stress \cite{Lutsko1988, Cormier2001} could also be merely a mathematical artifact rather than a reflection of the reality. Despite those problems, it offers many helpful insights to capture some essentials of the microscale stress field. First, it uses the local balances of linear momentum as the starting point, consistent with continuum mechanics, suggesting its potential compatibility with the stress determined by continuum mechanics. Regardless of the complicated derivation, the result is a summation of interactions between each interacting pairs, together with characteristic tensors in terms of distance vectors. This suggests a reverse approach: if we assume that along the line of interaction of an interacting pair the stress, when simply considered as force per unit area, over a certain characteristic area is dominated by their interatomic force, then we will be able to relate the stress directly to the atomic interactions by the Cauchy's relation, skipping the complications brought by the differential equations. On the other hand, for the majority of atoms there are many interacting neighbors, which suggests that we can use the Cauchy's relation to determine the resultant traction in many directions. Since the distance vectors can be obtained directly from the atomic positions, this then becomes an over-determined problem which can be solved by the least-square method. Of course, the results obtained would in general not be a very accurate one at every position, as the stress is varying from point to point. However, it must be kept in mind that we are not seeking for an accurate analytical expression of the microscale stress at any spatial position, but instead seeking for an approximation for a small region surrounding each individual atoms, i.e. the stress field in a small region around an individual atom is assumed to be constant. The entire stress field can then be interpolated from those point values by shape functions.

Comparing to the attention to the microscale stress, discussions on the microscale strain are rare in literature. The strain is usually considered as a given variable and used together with the microscale stress to obtain local elastic constants \cite{Lutsko1988, Cormier2001}. However, in continuum mechanics, strain itself can be used as an indicator of plastic deformation \cite{Simo1998} which, when translated into the microscale language, is the onset of dislocations. On the other hand, to estimate the microscale strain is equivalent to estimating the deformation gradient, since all kinds of strain tensors can be obtained from the deformation gradient \cite{Wriggers2009}. In the early date, Born \cite{Born1954} related the deformation gradient to the atomic positions in the lattice by the so called Cauchy-Born rule, under the assumption that the entire lattice is under uniform deformation. This has been later used as the foundation for the well known multiscale method named as the Quasicontinuum method \cite{Tadmor1996a, Miller2002}. If we imagine some virtual bonds exist between interacting pairs, then the assumption of homogeneous deformation means all the bonds are under the same deformation. For the microscale model where the particle methods must be applied, this assumption does not hold, i.e. the bond can be under different deformations. However, this does not mean that we have to abandon the Cauchy-Born rule completely. We can still apply it, but only to each bond. If, instead of seeking analytical expressions for the microscale strain, we want to obtain a reasonable engineering approximation for the microscale strain at individual atoms similar to what we hope for the stress tensor, then we can assume an averaged deformation gradient for each individual atom, and then apply the Cauchy-Born rule inversely, i.e. obtaining the deformation gradient from the undeformed and deformed distance vectors of each bond. As shown in the later section, as long as there are more than two linearly independent interacting bonds, such averaged deformation gradient can be estimated by the least-square method, just as the stress tensor. And the obtained deformation gradient can be further used to calculate the strain tensors at the positions of individual atoms.

In this short contribution, we will first derive the microscale strain tensor estimated by the least-square method, followed by the stress tensor. Then we show the results from the numerical verifications, where we compare the estimated microscale stress and strain tensors with the reference values. The numerical test is conducted on lattice model under simple loading conditions: uniform deformation and tensile test. Excellent matches between the least-square estimations and the reference values are obtained for the simple loading cases. After that, we present the results for more complicated loading conditions, including the stress field around a microscale crack, the stress field around an unit edge dislocation and the stress field under an indentation surface. In this contribution, we focus on the discussion of the method itself and the material parameters are of no interest, therefore the results of estimated stress field cannot be numerically compared to existing data. Nevertheless, the distribution pattern of the estimated stress shows good resemblance to the existing results in the literature. Limitations of our method are also discussed at the end.

\section{Least-Square Estimation of the Miroscale Strain Tensor}
\begin{figure}
  \centering
  \includegraphics[width = .4\textwidth]{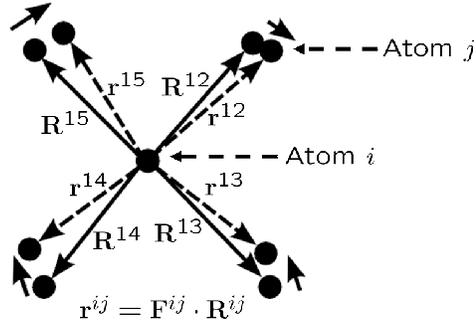}
  \caption{Deformed Bonds: distance vectors between atom $i$ and its neighbors, $j$, before and after deformation, related by the local deformation gradient $\mathbf{F}^{ij}$}
  \label{fig:Fij_dem}
\end{figure}
For atomic systems with only pair potentials, the interaction between any two atoms is independent from the others and it can be treated as virtual bond between these two atoms. Local deformation gradient can then be assigned to each bond, as shown in Fig.~\ref{fig:Fij_dem} and the deformed distance vector $\mathbf{r}^{ij}$ from arbitrary atom $i$ to its neighbor, atom $j$, can be related to their undeformed distance vector $\mathbf{R}^{ij}$ by applying the Cauchy-Born rule \cite{Tadmor1996a, Born1954},
\begin{equation}
  \mathbf{r}^{ij} = \mathbf{F}^{ij}\cdot\mathbf{R}^{ij}.
  \label{eq:cb_rule}
\end{equation}
To avoid confusion in their directions, we define the distance vectors here explicitly as
\begin{eqnarray}
    \mathbf{r}^{ij} & = \mathbf{x}^j - \mathbf{x}^i\\
    \mathbf{R}^{ij} & = \mathbf{X}^j - \mathbf{X}^i,
  \label{eq:def_r_R}
\end{eqnarray}
where $\mathbf{x}^i$ is the deformed position of atom $i$ and $\mathbf{X}^i$ its original position. All interaction neighbors of atom $i$ are indexed by $j$. We consider the interpolated deformation gradient at atom $i$ as an averaged value, from where we obtain the approximated distance vector after deformation. Since $i$ is arbitrary, without loss of generality we can drop the index $i$ from the notations for simplicity.
\begin{equation}
  \tilde{\mathbf{r}}^j = \mathbf{F}\cdot\mathbf{R}^j
  \label{eq:r_estimate}
\end{equation}
One can rewrite the right hand side of equation \eref{eq:r_estimate} as a mapping of the components of the deformation gradient as
\begin{equation}
\underbrace{\left[
  \begin{array}{ccc}
    {\mathbf{R}^j}^T & \mathbf{0} & \mathbf{0} \\
    \mathbf{0} & {\mathbf{R}^j}^T & \mathbf{0} \\
    \mathbf{0} & \mathbf{0} & {\mathbf{R}^j}^T \\
  \end{array}
\right]}_{\displaystyle\mathbf{D}^j}\hat{\mathbf{F}} = \tilde{\mathbf{r}}^j.
\label{eq:map_F}
\end{equation}
where $\hat{\mathbf{F}}$ is the Voigt notation of $\mathbf{F}$, defined as
\begin{equation}
  \hat{\mathbf{F}} = \left[F_{11}, F_{12}, F_{13}, F_{21}, F_{22}, F_{23}, F_{31}, F_{32}, F_{33}\right]^T.
  \label{eq:F_vec}
\end{equation}
The error of estimation of $\mathbf{r}^j$ is then
\begin{eqnarray}
        \Delta\mathbf{r}^j & = \tilde{\mathbf{r}}^j - \mathbf{r}^j\\
                    & = \mathbf{D}^j\hat{\mathbf{F}} - \mathbf{r}^j.
  \label{eq:err_bond_j}
\end{eqnarray}
The best estimation of $\mathbf{F}$ should minimize the error norm summed over $N_p$ interaction neighbors of atom $i$.
\begin{eqnarray}
  \Pi_F & =  \sum_{j=1}^{N_p}\left(\Delta\mathbf{r}^j\right)^T\Delta\mathbf{r}^j\\
  & = \sum_{j=1}^{N_p}\left(\mathbf{D}^j\hat{\mathbf{F}} - \mathbf{r}^j\right)^T\left(\mathbf{D}^j\hat{\mathbf{F}} - \mathbf{r}^j\right)
  \label{eq:err_norm_F}
\end{eqnarray}
Here we apply the standard linear least-square method, i.e. taking derivatives of equation \eref{eq:err_norm_F} and equaling it to zero. It yields a linear system,
\begin{equation}
  \sum_{j=1}^{N_p}\left({\mathbf{D}^j}^T\mathbf{D}^j\right)\hat{\mathbf{F}} - \sum_{j=1}^{N_p}\left({\mathbf{D}^j}^T\mathbf{r}^j\right) = \mathbf{0}.
  \label{eq:LS-F1}
\end{equation}
By performing some matrix manipulation, the best estimation of $\hat{\mathbf{F}}$ can be obtained by
\begin{equation}
  \hat{\mathbf{F}} = {\underbrace{\left(\mathbf{D}^T\mathbf{D}\right)}_{\displaystyle\mathbf{M}_F}}^{-1}\mathbf{D}^T\mathbf{b}_F,
  \label{eq:LS-F2}
\end{equation}
where
\begin{eqnarray}
  \mathbf{D} & = \left[\mathbf{D}_1^T,\mathbf{D}_2^T,\ldots,\mathbf{D}_{N_p}^T\right]^T\\
  \mathbf{b}_F & = \left[\mathbf{r}_1^T,\mathbf{r}_2^T,\ldots,\mathbf{r}_{N_p}^T\right]^T.
  \label{eq:def_DF}
\end{eqnarray}
Once $\mathbf{F}$ is recovered, the strain tensor can then be easily computed from
\begin{equation}
  \mathbf{E} = \frac{1}{2}\left(\mathbf{F}^T\mathbf{F}- \mathbf{I}\right),
  \label{eq:eps_def}
\end{equation}
where $\mathbf{I}$ is the 3-by-3 identity matrix. Notice that the matrix $\mathbf{M}_F$ in equation \eref{eq:LS-F2} depends only on the original lattice structure and has to be constructed only once, as far as no moving dislocation appears. On the other hand, to ensure that it is invertible, atom $i$ must have at least three interacting neighbors, whose distance vectors are linearly independent.
\section{Least-Square Estimation of the Miroscale Stress Tensor}
\begin{figure}
  \centering
  \includegraphics[width = .4\textwidth]{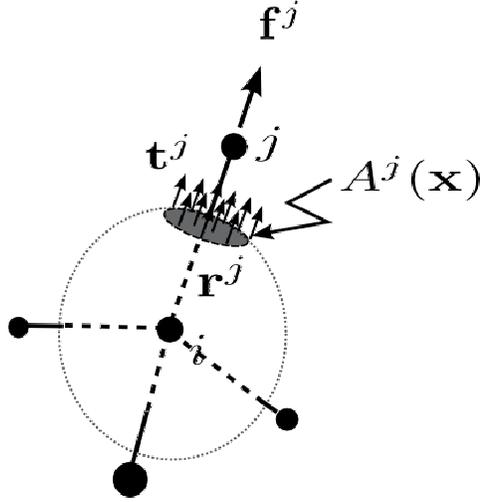}
  \caption{Equivalent traction along the direction of interaction}
  \label{fig:sig_A_cut}
\end{figure}
The stress estimation is a combination of the linear least-square method and the Cauchy's relation \cite{Heinbockel2002},
\begin{equation}
  \mathbf{t} = \boldsymbol{\sigma}\cdot\mathbf{n},
  \label{eq:cauchy_relation}
\end{equation}
where $\mathbf{t}$ is the traction vector associated to the Cauchy stress tensor in a cutting plane with unit normal vector $\mathbf{n}$. In static case, the atoms are always in equilibrium. For each atom $i$, we have
\begin{equation}
  \sum_{j=1}^{N_p}\mathbf{f}^j = \mathbf{0},
  \label{eq:fi_balance}
\end{equation}
where $\mathbf{f}^j$ is the force obtained from the interatomic potential between atom $i$ and $j$. Therefore, if we consider a special volume surrounding atom $i$ with its surface cutting through the bond $\mathbf{r}^j$, then the total external force at the intersection point can be equivalent to the interatomic force between atoms $i$ and $j$, as shown in Fig.~\ref{fig:sig_A_cut}, where the traction near by can be approximated by
\begin{equation}
  \mathbf{t}^j = \frac{\mathbf{f}^j}{A^j(\mathbf{x})}.
  \label{eq:t_ij}
\end{equation}
We define here $A(\mathbf{x})$ as some characteristic area, evaluated at position $\mathbf{x}$, for which equation \eref{eq:t_ij} holds. Together with equation \eref{eq:fi_balance}, one obtains
\begin{equation}
  \boldsymbol{\sigma}(\mathbf{x})\cdot\frac{\mathbf{r}^j}{\left|\mathbf{r}^j\right|} = \frac{\mathbf{f}^j}{A^j(\mathbf{x})},
  \label{eq:sig_rj}
\end{equation}
where $\left|\mathbf{r}^j\right|$ is the length of $\mathbf{r}^j$. For each atom $i$, we only estimate an averaged stress tensor within its surrounding space and therefore the unknown area $A^j(\mathbf{x})$ can be replaced by a constant value $A_c^j$ for each bond $\mathbf{r}^j$. The estimated stress tensor results in an approximated traction $\tilde{\mathbf{t}}^j$ for each bond with an error of
\begin{eqnarray}
    \Delta\mathbf{t}^j & = \tilde{\mathbf{t}^j} - \mathbf{t}^j \\
        & = \mathbf{d}^j\hat{\boldsymbol{\sigma}} - \frac{\mathbf{f}^j}{A_c^j},
  \label{eq:dt_ij}
\end{eqnarray}
where
\begin{eqnarray}
    \hat{\boldsymbol{\sigma}} & = \left[\sigma_{11}, \sigma_{12}, \sigma_{13}, \sigma_{21}, \sigma_{22}, \sigma_{23}, \sigma_{31}, \sigma_{32}, \sigma_{33}\right]^T\\
    \mathbf{d}^j & = \left[\begin{array}{ccc}
    {\mathbf{r}^j}^T & \mathbf{0} & \mathbf{0} \\
    \mathbf{0} & {\mathbf{r}^j}^T & \mathbf{0} \\
    \mathbf{0} & \mathbf{0} & {\mathbf{r}^j}^T
  \end{array}\right].
  \label{eq:sig_vec}
\end{eqnarray}
Then we define the total error norm for the traction as
\begin{eqnarray}
  \Pi^0_{\sigma} & = \sum_{j=1}^{N_p}\left(\Delta\mathbf{t}^j\right)^T\Delta\mathbf{t}^j \\
  & = \sum_{j=1}^{N_p}\left(\mathbf{d}^j\hat{\boldsymbol{\sigma}} - \frac{\left|\mathbf{r}^j\right|}{A_c^j}\mathbf{f}^j\right)^T\left(\mathbf{d}^j\hat{\boldsymbol{\sigma}} - \frac{\left|\mathbf{r}^j\right|}{A_c^j}\mathbf{f}^j\right),
  \label{eq:err_norm_sig}
\end{eqnarray}
whose compact form can be written as
\begin{equation}
  \Pi^0_{\sigma} = \left(\mathbf{d}\hat{\boldsymbol{\sigma}}-\mathbf{b}_{\sigma}\right)^T\left(\mathbf{d}\hat{\boldsymbol{\sigma}}-\mathbf{b}_{\sigma}\right),
  \label{eq:err_sig_compact}
\end{equation}
with
\begin{eqnarray}
  \mathbf{d} & = \left[\mathbf{d}_1^T,\mathbf{d}_2^T,\ldots,\mathbf{d}_{N_p}^T\right]^T\\
  \mathbf{b}_{\sigma} & = \left[\frac{\left|\mathbf{r}^1\right|}{A_c^1}{\mathbf{f}^1}^T, \frac{\left|\mathbf{r}^2\right|}{A_c^2}{\mathbf{f}^2}^T,\ldots,\frac{\left|\mathbf{r}^{N_p}\right|}{A_c^{N_p}}{\mathbf{f}^{N_p}}^T\right]^T.
  \label{eq:def_Dsig}
\end{eqnarray}
The interatomic potential for pair-interactions can written be in terms of the distance between the two atoms, i.e. $\Phi^j = \Phi^j(\left|\mathbf{r}^j\right|)$, and therefore the force can be derived as
\begin{equation}
  \mathbf{f}^j = \frac{{\Phi^j}'}{\left|\mathbf{r}^j\right|}\mathbf{r}^j.
  \label{eq:fij}
\end{equation}
By substituting equation \eref{eq:fij} into equation \eref{eq:def_Dsig}, $\mathbf{b}_{\sigma}$ can be simplified as
\begin{equation}
  \mathbf{b}_{\sigma} = \left[\frac{{\Phi^1}'}{A_c^1}{\mathbf{r}^1}^T, \frac{{\Phi^2}'}{A_c^2}{\mathbf{r}^2}^T,\ldots,\frac{{\Phi^{N_p}}'}{A_c^{N_p}}{\mathbf{r}^{N_p}}^T\right]^T.
  \label{eq:b_sig_simp}
\end{equation}
Meanwhile, the conservation of angular momentum for non-polar continua requires $\boldsymbol{\sigma}$ to be symmetric, i. e.
\begin{equation}
  \boldsymbol{\sigma} = {\boldsymbol{\sigma}}^T.
  \label{eq:sig_sym}
\end{equation}
When written in terms of elements it yields
\begin{equation}
  \sigma_{12} = \sigma_{21}, \sigma_{13} = \sigma_{31}, \sigma_{23} = \sigma_{32},
  \label{eq:sig_sym_elem}
\end{equation}
which can be converted to matrix form,
\begin{equation}
  \underbrace{\left[
     \begin{array}{ccccccccc}
       0 & 1 & 0 & -1 & 0 & 0 & 0 & 0 & 0 \\
       0 & 0 & 1 & 0 & 0 & 0 & -1 & 0 & 0 \\
       0 & 0 & 0 & 0 & 0 & 1 & 0 & -1 & 0 \\
     \end{array}
   \right]}_{\displaystyle\mathbf{A}}
  \hat{\boldsymbol{\sigma}} = \mathbf{0}
  \label{eq:sig_sym_mat}
\end{equation}
The symmetric constraint defined by equation \eref{eq:sig_sym_mat} can be imposed on the estimation by adding a penalty term to equation \eref{eq:err_sig_compact}, as
\begin{equation}
  \Pi_{\sigma} = \left(\mathbf{d}\hat{\boldsymbol{\sigma}}-\mathbf{b}_{\sigma}\right)^T\left(\mathbf{d}\hat{\boldsymbol{\sigma}}-\mathbf{b}_{\sigma}\right) + \lambda\left(\mathbf{A}\hat{\boldsymbol{\sigma}}\right)^T\left(\mathbf{A}\hat{\boldsymbol{\sigma}}\right),
  \label{eq:err_sig_mod}
\end{equation}
where $\lambda$ is the penalty factor. By minimizing equation \eref{eq:err_sig_mod} one can obtain the best estimation of the stress tensor $\hat{\boldsymbol{\sigma}}$ for atom $i$.
\begin{equation}
  \hat{\boldsymbol{\sigma}} = {\underbrace{\left({\mathbf{d}}^T\mathbf{d} + \lambda\mathbf{A}^T\mathbf{A}\right)}_{\displaystyle\mathbf{M}_{\sigma}}}^{-1}\mathbf{d}^T\mathbf{b}_{\sigma}.
  \label{eq:LS-sig}
\end{equation}
The matrix form of $\hat{\boldsymbol{\sigma}}$ can be recovered by equation \eref{eq:sig_vec}. To guarantee $\mathbf{M}_{\sigma}$ is invertible, at minimum 3 neighbors with linearly independent distance vectors are required for each atom, the same as the condition for the strain estimation.

So far we have not been able to derive an analytical form for the characteristic area $A_c^j$ used in equation \eref{eq:err_sig_compact}. By computational studies we found that the formula
\begin{equation}
  A_c^j = \frac{4\pi\left(\frac{1}{2}\left|\mathbf{R}^j\right|\right)^2}{N_p}
  \label{eq:A_c}
\end{equation}
provides the estimated $\mathbf{\sigma}$ with averaged values that are closest to the macroscale values. $\left|\mathbf{R}^j\right|$ is length of the undeformed distance vector between atom $i$ and $j$. It can be interpolated as a fraction of area of a spherical surface which is centered at atom $i$ and crosses the middle of the undeformed bond $\mathbf{R}^j$. And the fraction is determined by the number of interaction neighbors of atom $i$.

\section{Numerical Verification}
An atomic lattice model is constructed for conducting the numerical verifications. The interpolated microscale strain and stress fields are then compared with reference values.

\begin{figure}
  \centering
  \includegraphics[width = .8\textwidth]{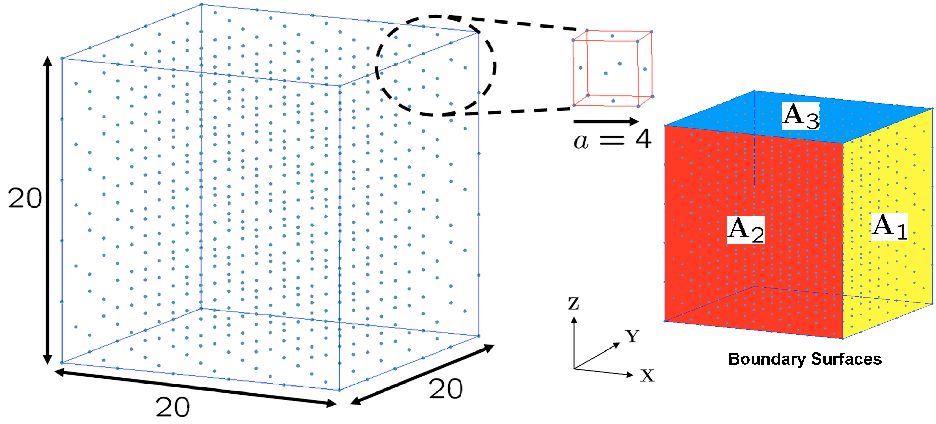}
  \caption{Lattice model used for numerical tests and boundary surfaces defined for equation \eref{eq:ref_sig}}
  \label{fig:lat_model}
\end{figure}
The lattice is a face-centered Bravais lattice with a lattice constant of 4, as illustrated in Fig.~\ref{fig:lat_model}. Lennard-Jones potential is applied, as
\begin{equation}
  \Phi^{ij} = \frac{B}{\left|\mathbf{r}^{ij}\right|^{12}} - \frac{A}{\left|\mathbf{r}^{ij}\right|^6},
  \label{eq:lj}
\end{equation}
where $A$ is chosen to be $1.0239\times10^8$ and $B$ to be $2.6211\times10^{10}$. The equilibrium distance is $2.8284$ and the cutoff radius is $3$ so that only immediate neighbors are interacting with each other. Since here the material properties are of no interest in the test, all parameters are therefore dimensionless.

In our numerical tests, uniform deformation is applied to the boundary atoms by
\begin{equation}
  \mathbf{u}_{bc} = \underbrace{\left(\mathbf{F} - \mathbf{I}\right)}_{\displaystyle\mathbf{H}}\mathbf{X}_{bc}.
  \label{eq:u_bc1}
\end{equation}
When the static equilibrium is reached the entire lattice will be under the same deformation gradient \cite{Born1954}, and the reference strain field can then be obtained by equation \eref{eq:eps_def}. On the other hand, the reference stress field is obtained by
\begin{equation}
  \boldsymbol{\sigma}_{ij}^{ref} = \frac{1}{A_i}\sum_k\mathbf{f}_j^{i, k},
  \label{eq:ref_sig}
\end{equation}
where $A_i$ is the area of the boundary surface $i$ defined in Fig.~\ref{fig:lat_model}, and $\mathbf{f}_j^{i, k}$ is the $jth$ component of constraint force applied on atom $k$ on the boundary surface $i$. The surface change due to the deformation is ignored here, due to the small deformation applied.
\begin{figure}
  \centering
  \includegraphics[width = .2\textwidth]{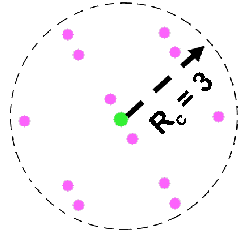}
  \caption{Reference lattice for calculating QC stress field: each atom has 12 neighbors when at the cutoff radius of 3, for a FCC lattice structure with lattice constant $a = 4$.}
  \label{fig:ref_lat}
\end{figure}

On the other hand, in order to investigate the possibility of applying our interpolation to multiscale method, we also compared our stress interpolation with the Quasicontinuum stress (QC) field \cite{Tadmor1996a}. To obtain the QC stress field, a simple reference lattice can be constructed as shown in Fig.~\ref{fig:ref_lat} whose radius is the same as the cutoff radius. The atoms at the center will be considered as the representative atom. Under uniform deformation, the 1st Piola-Kirchhoff stress tensor can be derived as \cite{Tadmor1996a}
\begin{equation}
  \mathbf{P} = \frac{1}{2V}\sum_{j=1}^{N_p}{\Phi^j}'\frac{1}{\left|\mathbf{r}^j\right|}\mathbf{r}^j\otimes\mathbf{R}^j,
  \label{eq:PK1}
\end{equation}
where $j$ is the index of the interaction neighbors of the representative atom and $V$ is the volume of the sphere surrounding the representative atom, at half of the cutoff radius. And the Cauchy stress tensor \cite{Wriggers2009} can then be obtained by
\begin{equation}
  \boldsymbol{\sigma} = J^{-1}\mathbf{P}\mathbf{F},
  \label{eq:cauchy_pk1}
\end{equation}
where $J$ is the determinant of the deformation gradient $\mathbf{F}$. The QC stress field represents the stress field for a crystal lattice of infinite size under uniform deformation.
\begin{figure}
  \centering
  \subfloat[Recovered strain, $E_{33}$]{\includegraphics[width = .5\textwidth]{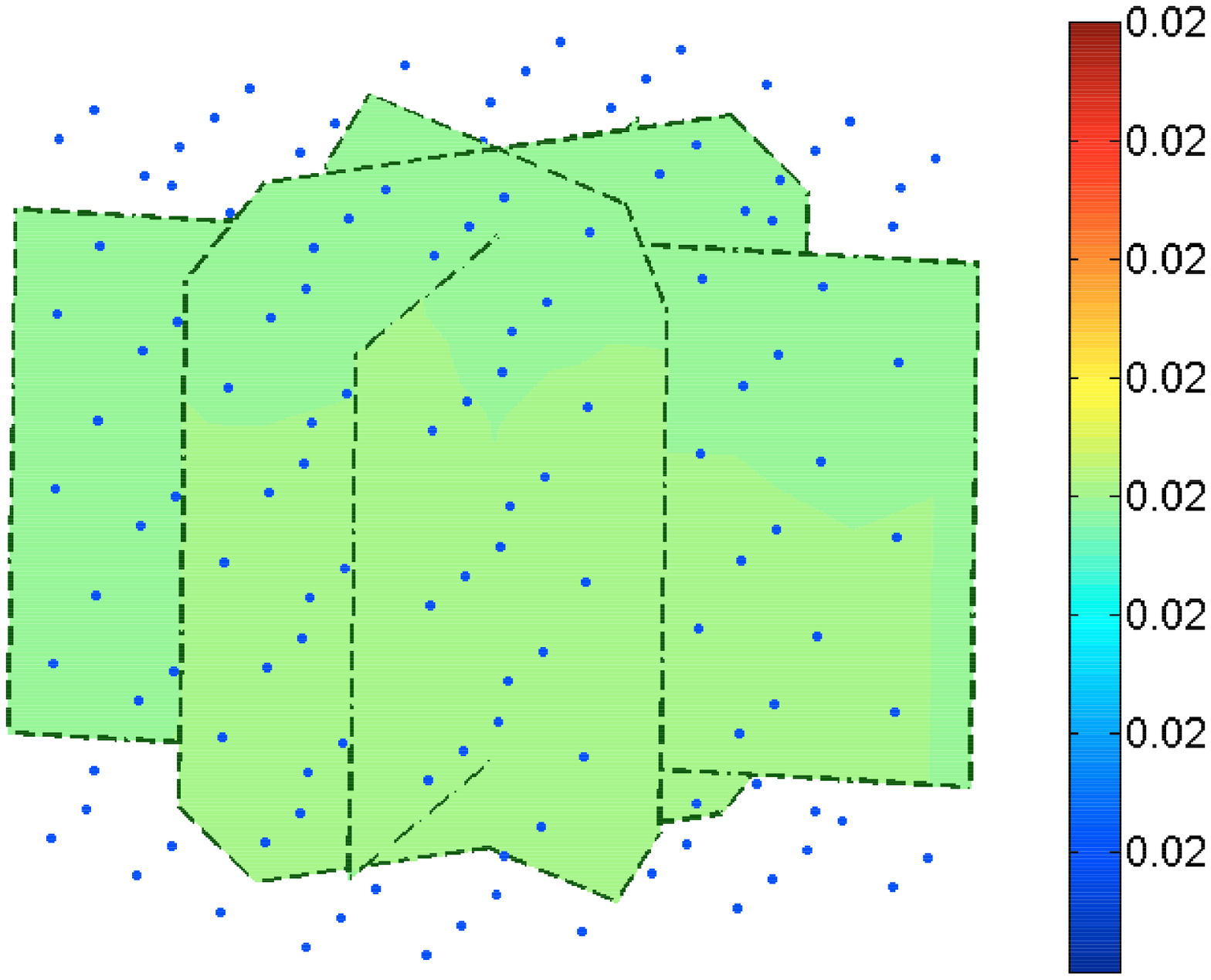}\label{fig:eps_noavg}}
  \subfloat[Recovered stress, $\sigma_{33}$]{\includegraphics[width = .5\textwidth]{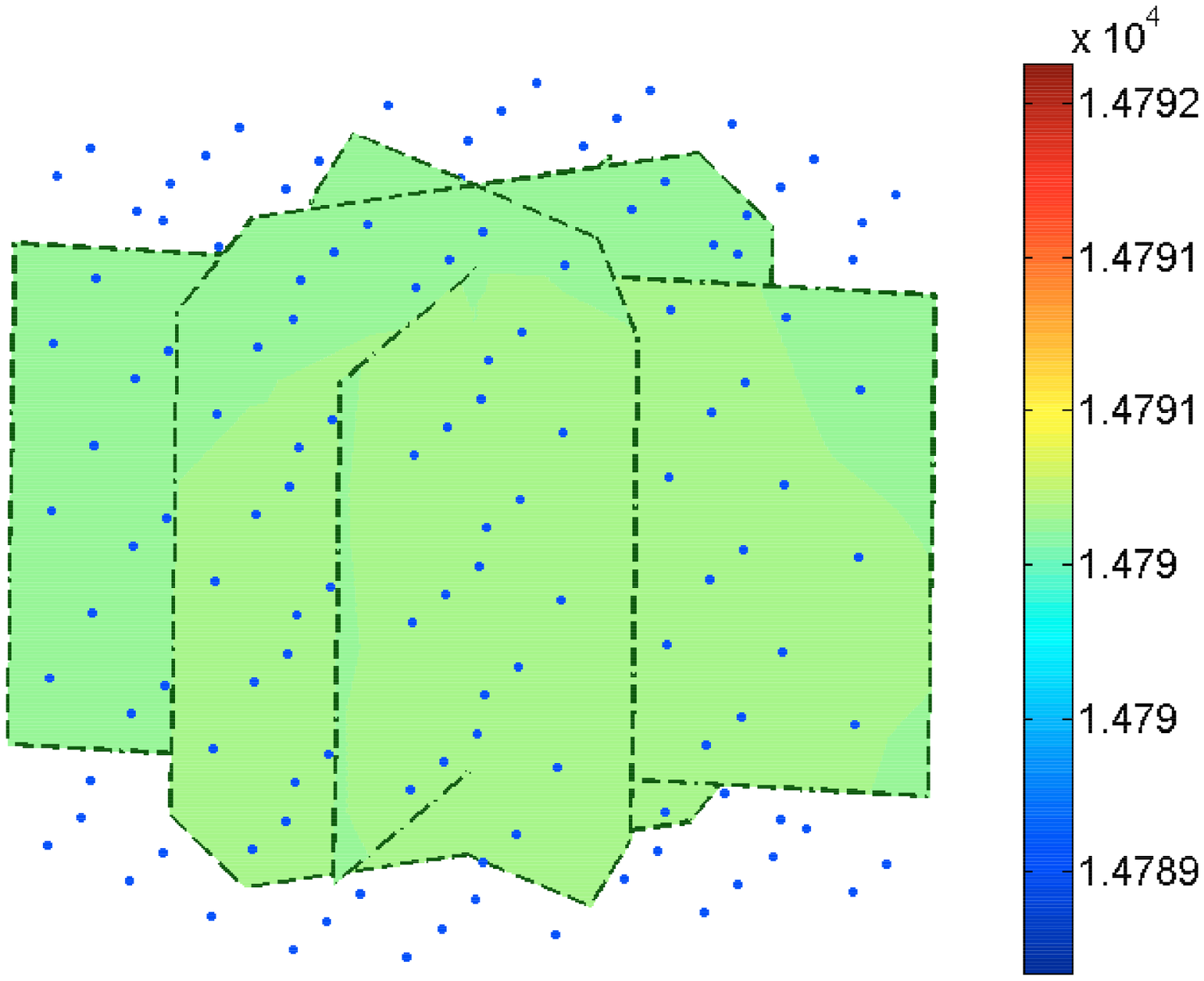}\label{fig:sig_noavg}}
  \caption{Recovered stress and strain fields: (a) Recovered strain field, $E_{33}$, where the reference value is $0.0200$. (b) Recovered stress field (without averaging), $\sigma_{33}$, where the reference value is $12164$. Notice that the recovered stress and strain are also uniform as the reference values.}
\end{figure}
\begin{table}
\centering
 \caption{Statistics of recovered strain and stress field (without averaging)}
  \label{tab:eps_sig_no_avg}
\begin{tabular}{|l|llllll|}
  \hline
  & $\epsilon_{11}$ & $\epsilon_{22}$ & $\epsilon_{33}$ & $\epsilon_{12}$ & $\epsilon_{13}$ & $\epsilon_{23}$ \\\hline
  $\mathbf{\epsilon}^{ref}$ & 0.0100 & 0.0100 & 0.0200 & 0.0100 & -0.0080 & 0.0100 \\
  $\mathbf{\epsilon}^{mean}$ & 0.0100 & 0.0100 & 0.0200 & 0.0100 & -0.0080 & 0.0100 \\
  $Std(\mathbf{\epsilon})(10^{-14})$ & 0.14 & 0.18 & 0.17 & 0.10 & 0.10 & 0.11 \\\hline
    & $\sigma_{11}$ & $\sigma_{22}$ & $\sigma_{33}$ & $\sigma_{12}$ & $\sigma_{13}$ & $\sigma_{23}$ \\\hline
  $\mathbf{\sigma}^{ref}$ & 10143 & 9997 & 12164 & 4035 & -2860 & 3610 \\
  $\mathbf{\sigma}^{QC}$  & 9057  & 8939 & 10954 & 3643 & -2592 & 3286 \\
  $\mathbf{\sigma}^{mean}$ & 12401 & 12206 & 14790 & 4928 & -3438 & 4342 \\
  Err($\mathbf{\sigma}^{mean}$)($\%$) & 22.26 & 22.09 & 21.59 & 22.13 & 20.19 & 20.27 \\
  $Std(\mathbf{\sigma})(10^{-9})$ & 0.69 & 0.64 & 0.72 & 0.45 & 0.56 & 0.42 \\
  \hline
\end{tabular}
\end{table}

For the first test, combined uniform strain field is applied to the lattice,
\begin{equation}
  \mathbf{E} = \left[
                 \begin{array}{ccc}
                   0.010 & 0.010 & -0.008 \\
                   0.010 & 0.010 & 0.010 \\
                   -0.008 & 0.010 & 0.020 \\
                 \end{array}
               \right].
\end{equation}
The interpolation results for selected lattice planes are plotted in Fig.~\ref{fig:eps_noavg} and Fig.~\ref{fig:sig_noavg}, for the strain field and stress field accordingly, and the corresponding statistics are listed in Table.~\ref{tab:eps_sig_no_avg}. Surface atoms are excluded from the final results. The recovered microscale stress and strain fields are uniform, same as the reference values. The strain estimation shows almost perfect match with the reference values. The stress estimation on the other hand shows about $20\%$ error compared to the reference values, as listed in Table.~\ref{tab:eps_sig_no_avg}. This shift is caused by the surface atoms which are under unsymmetric interactions. We figured out that by combining the stress at each atom $i$ with its interacting neighbors $j$ by weighted summation, as
\begin{equation}
  \bar{\mathbf{\sigma}}^i = \frac{1}{2}\mathbf{\sigma}^i + \frac{1}{2N_p}\sum_{j=1}^{N_p}\mathbf{\sigma}^j,
  \label{eq:avg_sig}
\end{equation}
the estimation for the internal stress can be significantly improved. The recovered stress values after applying the averaging technique equation \eref{eq:avg_sig} are given in Table.~\ref{tab:sig_tab_yesavg}. One can observe that the internal stress field remains to be uniform while its accuracy is significantly improved, with the error reduced to the level of $2\%$. Moreover, on can observe that the stress field becomes much more closer to the QC stress field, which implies a promising application of the technique to the multiscale methods based on the QC framework.
\begin{table}
\centering
 \caption{Statistics of recovered stress field (averaged by equation \eref{eq:avg_sig})}
  \label{tab:sig_tab_yesavg}
\begin{tabular}{|l|llllll|}
  \hline
    & $\sigma_{11}$ & $\sigma_{22}$ & $\sigma_{33}$ & $\sigma_{12}$ & $\sigma_{13}$ & $\sigma_{23}$ \\\hline
  $\mathbf{\sigma}^{ref}$ & 10143 & 9997 & 12164 & 4035 & -2860 & 3610 \\
  $\mathbf{\sigma}^{QC}$  & 9057  & 8939 & 10954 & 3643 & -2592 & 3286 \\
  $\mathbf{\sigma}^{int}$ & 10335 & 10173 & 12326 & 4069 & -2839 & 3586 \\
  Err($\mathbf{\sigma}^{int}$)($\%$) & 1.86 & 1.76 & 1.33 & 0.85 & -0.76 & -0.68 \\
  $Std(\mathbf{\sigma})(10^{-9})$ & 0.35 & 0.32 & 0.36 & 0.22 & 0.28 & 0.21 \\
  \hline
\end{tabular}
\end{table}

\begin{figure}[htp]
  \centering
  \includegraphics[width = .9\textwidth]{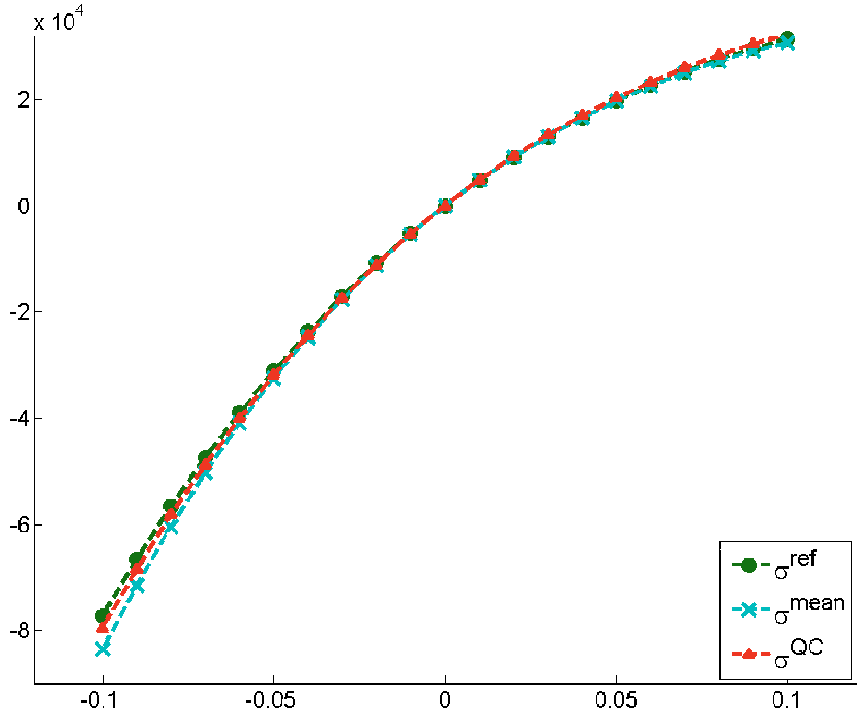}
  \caption{Stress vs. Strain for the tensile test: Stress-Strain Curves (Right)}
  \label{fig:sig_tensile}
\end{figure}

Moreover, to investigate the performance of our interpolation changing with the severity of the deformation a numerical tensile test is conducted on the lattice model. An uniform strain in the z-direction is applied,
\begin{equation}
  \mathbf{E} = \left[
                 \begin{array}{ccc}
                   0 & 0 & 0 \\
                   0 & 0 & 0 \\
                   0 & 0 & E_{33} \\
                 \end{array}
               \right]
\end{equation}
where $E_{33}$ varies from $-0.1$ to $0.1$ for $20$ loading steps. The recovered values of the normal stress in the loading direction are plotted in Fig.~\ref{fig:sig_tensile}, together with reference values. The averaging technique defined by equation \eref{eq:avg_sig} is applied during the recovery. From Fig.~\ref{fig:sig_tensile} one can observe that the recovered stress values matches very well with the reference value and the QC stress field during the entire test. It can be also noticed that the QC stress field shows better consistency with the reference values than the previous test where a complicated strain field was applied. The unsymmetric response between tension and compression comes from the strong repulsion of the interatomic potentials when atoms are pressed together.
\section{Examples of Complicated Loading Conditions}
\begin{figure}[htp]
  \centering
  \subfloat[Micro crack (Von Mises)]{\includegraphics[width = .5\textwidth]{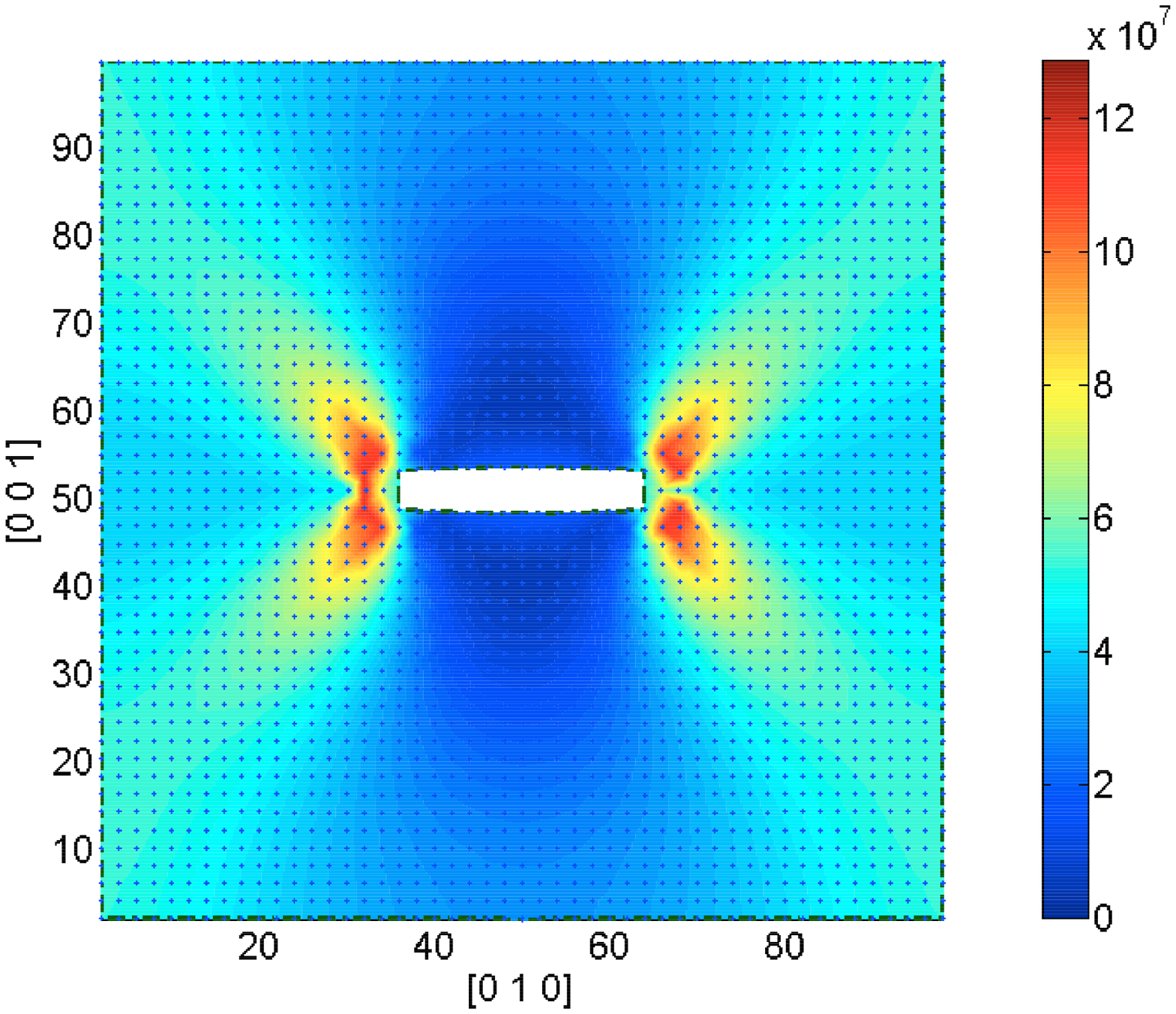}\label{fig:sig_crack}}
  \subfloat[Unit edge dislocation ($\sigma_{zz}$)]{\includegraphics[width = .5\textwidth]{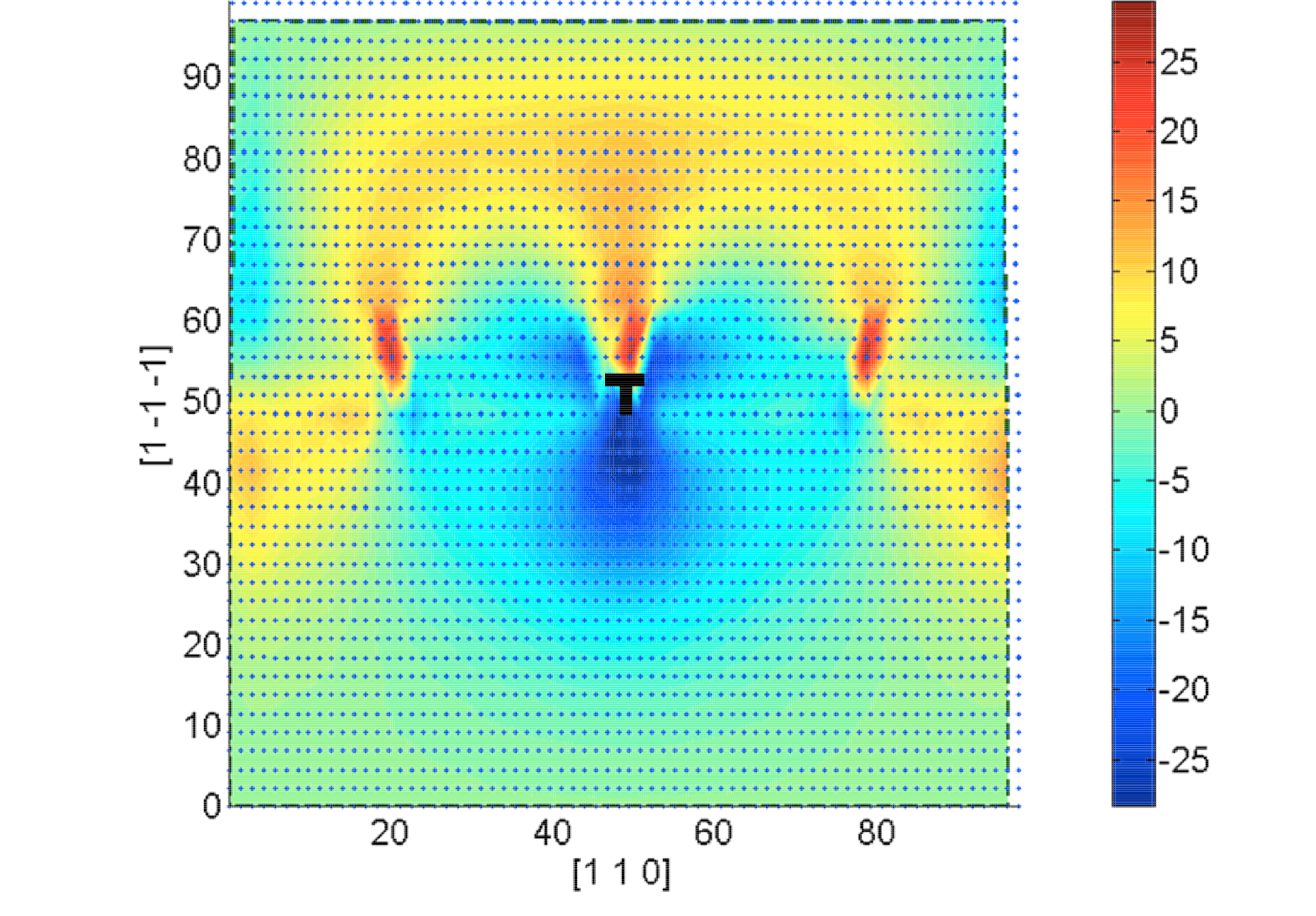}\label{fig:sig_disloc}}\\
  \subfloat[Indentation (Von Mises)]{\includegraphics[width = .5\textwidth]{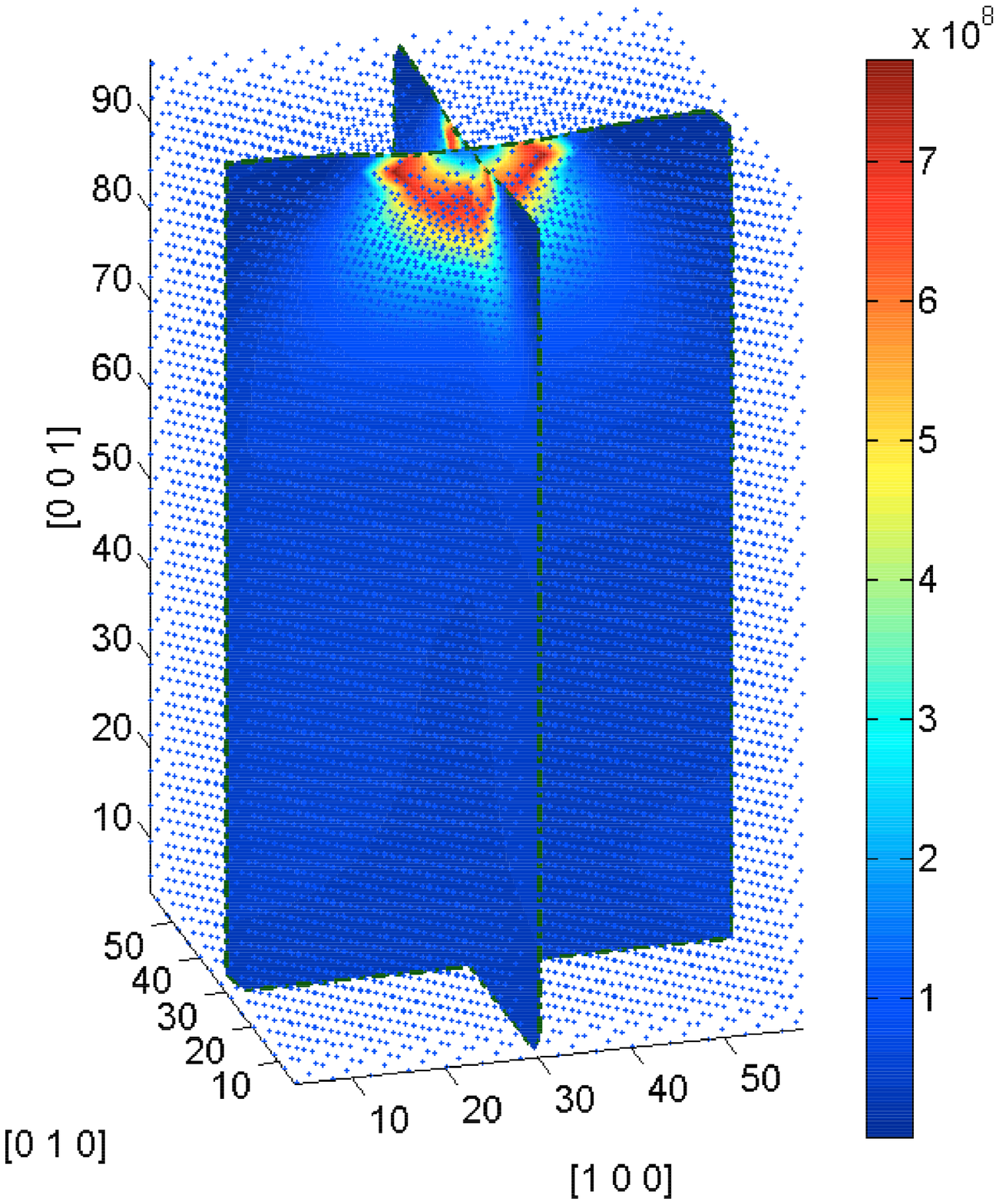}\label{fig:sig_indent}}
  \subfloat[Indentation with defects (Von Mises)]{\includegraphics[width = .5\textwidth]{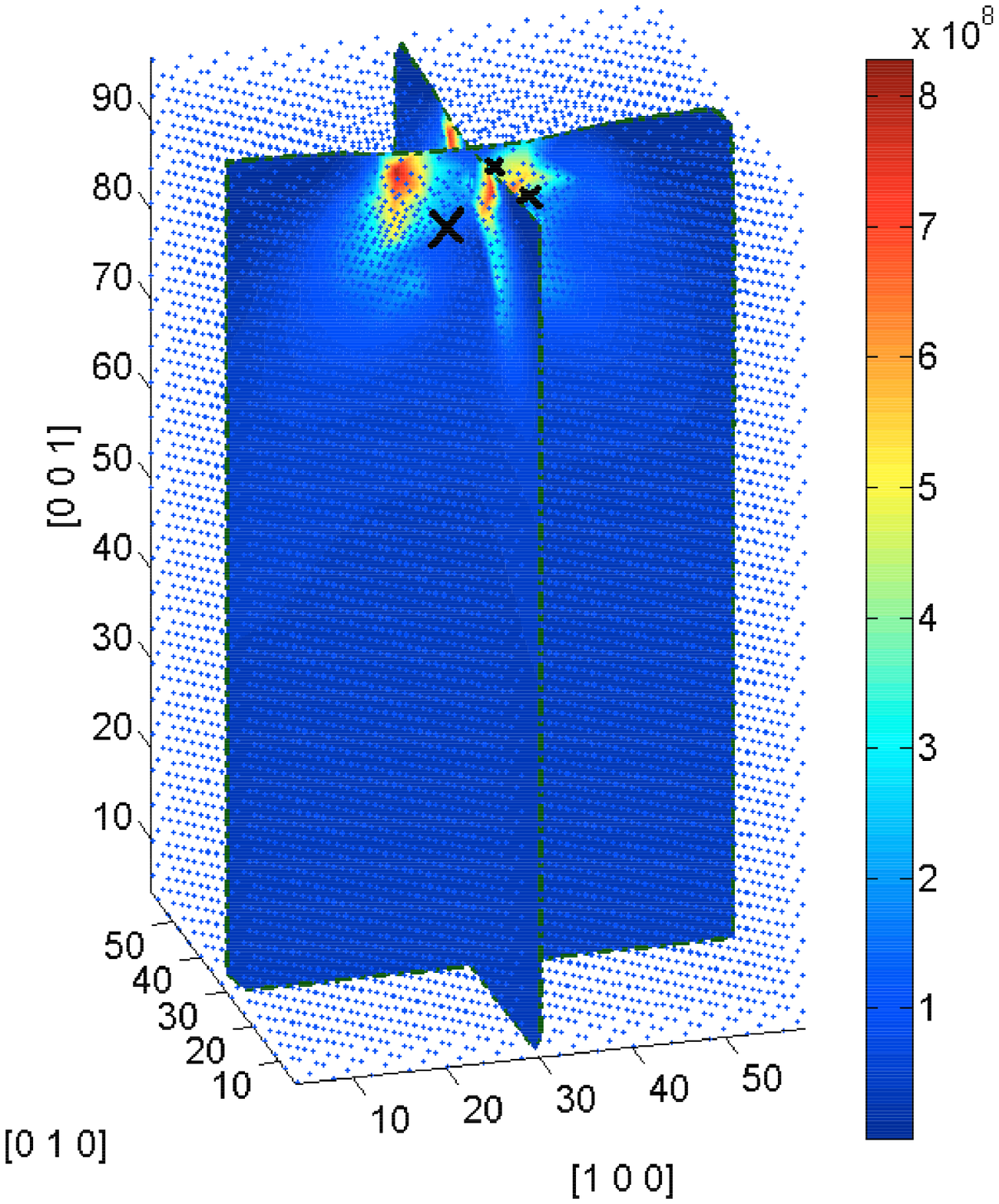}\label{fig:sig_indent_defkt}}
  \caption{Microscale stress fields for complicated loading conditions}
\end{figure}

After the verification for simple loading cases, we then applied the interpolation method to study some more practical problems. In this contribution we show the interpolated stress fields around a micro crack, around an unit edge dislocation in the FCC lattice and below an indentation surface. Figure.~\ref{fig:sig_crack} shows the Von Mises stress around a micro crack, generated by removing a single layer of atoms in the middle of lattice. The lattice is modeled in the shape of a thin plate with a thickness of 10, and is uniformly pulled in the $[0 0 1]$ direction. The figure shows that the highest stresses interpolated by our method occur at the corners of the crack and stretches in the diagonal directions, whereas the atoms right above and below the crack surface experience almost no stress. The places with high stress concentrations can be considered as potential positions for dislocations.

In Figure.~\ref{fig:sig_disloc}, an unit edge dislocation with a burger vector of $\frac{1}{2}[1 1 0]$ is created in the middle and the resultant stress field $\sigma_{zz}$ after relaxation is illustrated. Due to the limitation of computational power, only immediate neighbors are interacting with each other. The interpolated result indicates a symmetric distribution below and above the core, resembling the pattern predicted by the continuum theories \cite{Hull2001}. However, this symmetry is distorted by the boundary effects. On the other hand, the dislocation introduces a pair of symmetric concentrations far away from the dislocation, indicating a tendency of splitting during the relaxation. This could be an intrinsic behavior of the dislocation or could merely be an influence of the boundary effects.

At the end, examples of stress distribution under an indentation surface are shown in Figure.~\ref{fig:sig_indent} and Figure.~\ref{fig:sig_indent_defkt}, where a square indentor is pressed into the center of the surface. The figure shows the distribution of the Von Mises stress. In Figure.~\ref{fig:sig_indent} one can observe that when there is no defect, stress is concentrated at the corners and the bottom of the indentor. When there are defects (point vacancies) closely beneath the indentation surface, the stress distribution is notably changed. Stress is severely concentrated around the defects and the maximum value is significantly increased. On the other hand, under the combined effects of the indentation and the lattice defects, the stress in certain region is observed to be lowered.
\section{Conclusions and Discussion}
In this contribution we introduced a straightforward approach to interpolate the microscale strain and stress for particle systems. The numerical test shows very good consistency between the recovered strain field and the applied strain field for simple loading conditions. The stress estimation shows relatively large errors compared to the reference values, which can be effectively reduced by applying averaging techniques. On the other hand, we observe good consistency between the interpolated microscale stress field and the stress field obtained by the Quasicontinuum method, which indicates potential applications of the method in multiscale methods. The tensile test shows that the error between the interpolated stress field and the reference stress field stays at low level during the entire loading process, further justifying the applicability of our method. For complicated loading conditions, such as dislocations and indentations, we observed that the patterns of the estimated stress resembles the results from the literature. There are also some foreseen limitations associated with our method. First, there is no dynamic effects considered for the stress estimation, or in other words the influence of mass flow, if any, to the stress is not considered. Second, it is possible that the results given in this contribution only valid for particles systems under pair potentials, nevertheless, we believe similar procedures can be applied to systems with many-body interactions.
\section{Acknowledgements}
The authors gratefully acknowledge support from the Graduate Research Group 615 (GRK615), funded by the German Research Foundation (DFG).

\end{document}